\newcommand{\lsim}
{\;\raisebox{-.3em}{$\stackrel{\displaystyle <}{\sim}$}\;}

\documentclass{ws-rv961x669}
\usepackage{ws-rv-van}     
\usepackage{ws-rv-thm}     
\usepackage{subfigure}     
\makeindex

\begin{document}

\chapter[Colour Octet Extension of 2HDM]{Colour Octet Extension of 2HDM}\label{ra_ch1}

\author[]{German Valencia\footnote{German.Valencia@monash.edu}\footnote{On leave from Department of Physics, Iowa State University, Ames, IA 50011.}}

\address{School of Physics and Astronomy\\ Monash University, Melbourne, Australia. \\}

\begin{abstract}
In this talk we consider some aspects of the Manohar-Wise extension of the SM with a colour-octet electroweak-doublet scalar applied to 2HDM.  We present theoretical constraints on the parameters of this extension to both the SM and the 2HDM and discuss related phenomenology at LHC.

\end{abstract}


\body


\section{Introduction}\label{ra_sec1}

Now that the Higgs boson has been found \cite{Aad:2012tfa,Chatrchyan:2012ufa} everyone is asking whether it 
is really THE Higgs. Alternatively, the question is whether there any more scalars, and we got a hint last
December that there may be one at 750 GeV \cite{CMS:2015dxe,ATLAS750}. 
Many extensions of the scalar sector of the SM have been studied: two higgs doublet models, extra singlets, triplets ...  but mostly colour singlets. 
There are known phenomenological constraints on these  extensions: 
triplets tend to run into trouble with the $\rho$ parameter; 
multiple doublets introduce FCNC, and so on. FCNC can be avoided in several ways: one can require 
each doublet to couple to same charge quarks (known as type II 2HDM) \cite{Glashow:1976nt}; or one can impose 
minimal flavour violation  \cite{Chivukula:1987py,D'Ambrosio:2002ex}, where all flavour breaking is due to Yukawa matrices.

Assuming that the new scalars are singlets under the flavour group, MFV allows only SU(2) doublets that are colour singlets or colour octets, and the addition of a colour octet to the SM was introduced by Manohar and Wise (MW)  \cite{Manohar:2006ga}.  Colour singlets are the usual case of the 2HDM \cite{reviews} whereas 
colour octets are another possibility that has received less attention. 
Other possibilities exist if the scalars transform under the flavour group \cite{Arnold:2009ay}. There are also many 
complete models that have colour octet scalars, such as unification with 
$SU(5)$ \cite{Perez:2016qbo}; with 
$SO(10)$ \cite{Bertolini:2013vta}, models where the Higgs boson is not elementary such as topcolour \cite{Hill:1991at}, technicolour \cite{Farhi:1980xs} models and many more.

New colour octet scalars can have a large effect on loop level Higgs decay, which is also the dominant production mode \cite{Manohar:2006ga}. 
One of first examples of NP ruled out by the Higgs observation was the fourth generation, in which 
gluon fusion production of the Higgs would be about 10 times larger than in the SM, due to the extra contributions from $t^\prime,b^\prime$ in the loop. Similar large effects can appear in Higgs physics in the presence of colour octet scalars but in this case the couplings that appear in Higgs production can have either sign and their magnitude is a free parameter so that they could, for example, cancel the new contributions from a fourth generation \cite{He:2011ti}. Other examples of their exotic phenomenology are that they can produce the Higgs with completely different Yukawa coupling for top \cite{He:2013tia}, or induce large CP violation in Higgs production \cite{He:2011ws}. Even though there are many studies of this type \cite{Burgess:2009wm,Carpenter:2011yj,Enkhbat:2011qz,He:2011ti,Dobrescu:2011aa,Bai:2011aa,Arnold:2011ra,He:2011ws,Cacciapaglia:2012wb,Dorsner:2012pp,Kribs:2012kz,Reece:2012gi,Cao:2013wqa,He:2013tla,He:2013tia,Cheng:2015lsa,Buttazzo:2014bka,He:2014xla,Yue:2014tya}, these colour-scalars are very hard to see directly at LHC.

\section{The model}

The most general renormalizable scalar potential for the SM plus MW is\footnote{We use a normalization of $\lambda$  with the conventional relation $\lambda=G_Fm_H^2/\sqrt{2}$. We use $\tilde\lambda_i$ to distinguish from $\lambda_i$ of the 2HDM.}
\begin{eqnarray}
V&=&{\lambda}\left(H^{\dagger i}H_i-\frac{v^2}{2}\right)^2+2m_s^2\ {\rm Tr}S^{\dagger i}S_i +\tilde{\lambda}_1\ H^{\dagger i}H_i\  {\rm Tr}S^{\dagger j}S_j +\tilde{\lambda}_2\ H^{\dagger i}H_j\  {\rm Tr}S^{\dagger j}S_i 
\nonumber \\
&+&\left( \tilde{\lambda}_3\ H^{\dagger i}H^{\dagger j}\  {\rm Tr}S_ iS_j +\tilde{\lambda}_4\ e^{i\phi_4}\ H^{\dagger i} {\rm Tr}S^{\dagger j}S_ jS_i +
\tilde{\lambda}_5\ e^{i\phi_5}\ H^{\dagger i} {\rm Tr}S^{\dagger j}S_ iS_j 
+{\rm ~H.c.}\right)\nonumber \\
&+& \tilde{\lambda}_6\  {\rm Tr}S^{\dagger i}S_ iS^{\dagger j} S_j +\ 
 \tilde{\lambda}_7\  {\rm Tr}S^{\dagger i}S_ jS^{\dagger j} S_i +\
  \tilde{\lambda}_8\  {\rm Tr}S^{\dagger i}S_ i\ {\rm Tr}S^{\dagger j} S_j
  \nonumber \\
  &+&  \tilde{\lambda}_9\  {\rm Tr}S^{\dagger i}S_ j\ {\rm Tr}S^{\dagger j} S_i
  +\  \tilde{\lambda}_{10}\  {\rm Tr} S_i
S_ j\ {\rm Tr}S^{\dagger i}S^{\dagger j}+\ 
\tilde{\lambda}_{11}\  {\rm Tr} S_iS_ jS^{\dagger j}S^{\dagger i}.
\label{potential}
\end{eqnarray}
where $v\sim 246$~GeV is the Higgs vev.

After symmetry breaking, the non-zero vev of the Higgs gives the physical Higgs scalar $h$  a mass $m^2_H = 2 \lambda v^2$ and it also splits  the octet scalar masses as,
\begin{eqnarray}
m^2_{S^{\pm}} =  m^2_S + \tilde\lambda_1 \frac{v^2}{4},&&
m^2_{S^{0}_{R,I}} =  m^2_S + \left(\tilde\lambda_1 + \tilde\lambda_2 \pm 2\tilde \lambda_3 \right) \frac{v^2}{4},
\end{eqnarray}
The parameters $m_S^2$, and $\tilde\lambda_{1,2,3}$ should be chosen such that the above squared masses remain positive. 

Our next step  is to extend the type I and type II two Higgs doublet models with a colour octet electroweak doublet  as in  MW.  The scalar potential is required to satisfy desirable properties: minimal flavour violation  and custodial symmetry and is constructed in steps as follows.
We start with the well known  2HDM potential for ($\Phi_1, \Phi_2)$   assuming CP conservation and a discrete symmetry $\Phi_1\to -\Phi_1$ that is only violated softly by dimension two terms \cite{reviews}.\footnote{A condition that is more restrictive than MFV.}
\begin{eqnarray}
V\left( \Phi_1, \Phi_2 \right)  &=& m_{11}^2 \Phi_1^\dag \Phi_1 + m_{22}^2 \Phi_2^\dag \Phi_2 - m_{12}^2 \left( \Phi_1^\dag \Phi_2 + \Phi_2^\dag \Phi_1 \right) \nonumber \\
&+& \frac{\lambda_1}{2} \left( \Phi_1^\dag \Phi_1 \right)^2 + \frac{\lambda_2}{2} \left( \Phi_2^\dag \Phi_2 \right)^2 
+ \lambda_3 \left( \Phi_1^\dag \Phi_1 \right) \left( \Phi_2^\dag \Phi_2 \right)  \nonumber\\
&+& \lambda_4 \left( \Phi_1^\dag \Phi_2 \right) \left( \Phi_2^\dag \Phi_1 \right) + \frac{\lambda_5}{2} \left[ \left( \Phi_1^\dag \Phi_2 \right)^2 + \left( \Phi_2^\dag \Phi_1 \right)^2 \right].
\label{thdmV}
\end{eqnarray}

Next we can add the most general, renormalizable potential that describes the couplings of the colour octet $S$ to the two colour singlets ($\Phi_1, \Phi_2)$ as well as the self interactions of the colour octet. This potential can be easily constructed by analogy with Eq.~\ref{potential}. The octet self interactions do not change, 
\begin{eqnarray}
V(S) &=& 2m_S^2 {\rm Tr}S^{\dag i}S_i + \mu_1 {\rm Tr}S^{\dag i}S_i S^{\dag j}S_j + \mu_2 {\rm Tr}
S^{\dag i}S_j S^{\dag j}S_i + \mu_3 {\rm Tr} S^{\dag i}S_i {\rm Tr}S^{\dag j} S_j\nonumber\\
& +& \mu_4 {\rm Tr}S^{\dag i}S_j {\rm Tr}S^{\dag j}S_i + \mu_5 {\rm Tr}S_i S_j{\rm Tr}
S^{\dag i}S^{\dag j} + \mu_6 {\rm Tr}S_i S_j S^{\dag j}S^{\dag i}.
\label{sselfV}
\end{eqnarray}
Interactions between one of the two colour singlets and the colour octet  mimic Eq.~\ref{potential},
\begin{eqnarray}
V\left( \Phi_1, S \right)  &= &\nu_1 \Phi_1^{\dag i}\Phi_{1i}{\rm Tr}S^{\dag j}S_j + \nu_2 \Phi_1^{\dag i}\Phi_{1j}
{\rm Tr}S^{\dag j}S_i\nonumber\\
& +& \left( \nu_3 \Phi_1^{\dag i}\Phi_1^{\dag j}{\rm Tr}S_i S_j + \nu_4 \Phi_1^{\dag i}{\rm Tr}
S^{\dag j}S_j S_i + \nu_5 \Phi_1^{\dag i}{\rm Tr}S^{\dag j}S_i S_j + {\rm h.c.} \right) \nonumber \\
V\left( \Phi_2, S \right)  &= &\omega_1 \Phi_2^{\dag i}\Phi_{2i}{\rm Tr}S^{\dag j}S_j + \omega_2 \Phi_2^{\dag i}\Phi_{2j}
{\rm Tr}S^{\dag j}S_i\nonumber\\
& +& \left( \omega_3 \Phi_2^{\dag i}\Phi_2^{\dag j}{\rm Tr}S_i S_j + \omega_4 \Phi_2^{\dag i}{\rm Tr}
S^{\dag j}S_j S_i + \omega_5 \Phi_2^{\dag i}{\rm Tr}S^{\dag j}S_i S_j + {\rm h.c.} \right) 
\label{phisV}
\end{eqnarray}
Lastly, we look at terms that include both $\Phi_1$ and $\Phi_2$ as well as $S$,\footnote{Note that these terms are allowed by MFV but not by the discrete symmetry commonly used to restrict the 2HDM potental.}
\begin{eqnarray}
V_{N}\left( \Phi_1, \Phi_2, S \right) = \kappa_1 \Phi_1^{\dag i}\Phi_{2i}{\rm Tr}S^{\dag j}S_j +
\kappa_2 \Phi_1^{\dag i}\Phi_{2j}{\rm Tr}S^{\dag j}S_i + \kappa_3 \Phi_1^{\dag i}\Phi_2^{\dag j}{\rm Tr}S_j S_i
+ {\rm h.c.}
\label{3fields}
\end{eqnarray}
In our notation the  $SU(2)$ indices $i, j$ are shown explicitly,  $S_i = T^A S_i^A$, and the trace is over colour indices.
The complete potential is thus,
\begin{equation}
V\left( \Phi_1, \Phi_2, S \right) = V\left( \Phi_1, \Phi_2 \right) + V (S) + V\left( \Phi_1, S \right) +
V\left( \Phi_2, S \right) + V_{N}\left( \Phi_1, \Phi_2, S \right).
\end{equation}

After symmetry breaking,  this potential yields the following scalar masses
\begin{eqnarray}
m_{H^\pm}^2 &=& \frac{2 m_{12}^2}{\sin2\beta}-\frac{\lambda_4 + \lambda_5}{2}v^2,\quad\quad m_A^2 \, =\, \frac{2 m_{12}^2}{\sin2\beta} - \lambda_5 v^2,\nonumber \\
m_h^2 &=& \frac{2 m_{12}^2}{\sin2\beta} \cos^2 (\beta - \alpha)  + v^2 \left( \lambda_1 \sin^2 \alpha \cos^2 \beta + \lambda_2 \cos^2 \alpha \sin^2 \beta - \frac{\lambda_{345}}{2}\sin2\alpha \sin2\beta \right),\nonumber \\
m_H^2 &=& \frac{2 m_{12}^2}{\sin2\beta} \sin^2 (\beta - \alpha) + v^2 \left( \lambda_1 \cos^2 \alpha \cos^2 \beta + \lambda_2 \sin^2 \alpha \sin^2 \beta + \frac{\lambda_{345}}{2}\sin2\alpha \sin2\beta \right),\nonumber \\
m_{12}^2 &=& \frac{ v^2 \left[ \left(  \lambda_1 \cos^2 \beta - \lambda_2 \sin^2 \beta \right) \tan2\alpha - \frac{\lambda_{345}}{2}\sin2\beta \right]}{2 \tan2\alpha \cot2\beta - 1}.
\label{masses}
\end{eqnarray}
where $\lambda_{345}=\lambda_3+\lambda_4+\lambda_5$, and $v^2=v_1^2+v_2^2$ with $v_{1,2}$ the vevs of $\Phi_{1,2}$ respectively.  Similarly, for the colour octet sector 
\begin{eqnarray}
m_{{S^ \pm }}^2 &=& m_S^2 + \frac{v^2}{4}\left( \nu_1 \cos^2 \beta + \omega_1 \sin^2 \beta +
\kappa_1 \sin2\beta \right),\nonumber \\
m_{S_R^0}^2 &=& m_S^2 + \frac{v^2}{4} \left[ \left( \nu_1 + \nu_2 + 2 \nu_3 \right) \cos^2 \beta + \left( \omega_1 + \omega_2 + 2 \omega_3 \right) \sin^2 \beta \right. \nonumber\\
&+& \left. \left( \kappa_1 + \kappa_2 + \kappa_3 \right) \sin2\beta \right], \nonumber \\
m_{S_I^0}^2 &=& m_S^2 + \frac{v^2}{4} \left[ \left( \nu_1 + \nu_2 - 2 \nu_3 \right) \cos^2 \beta + \left( \omega_1 + \omega_2 - 2 \omega_3 \right) \sin^2 \beta \right. \nonumber\\
&+& \left.  \left( \kappa_1 + \kappa_2 - \kappa_3 \right) \sin2\beta \right].
\label{masses2}
\end{eqnarray}

The Yukawa couplings, $L_Y = L_{Y1}\left( \Phi_1, \Phi_2 \right) + L_{Y2}\left( S \right)$, 
 in the flavour eigenstate basis  are
\begin{eqnarray}
&L_{Y1}\left( \Phi_1, \Phi_2 \right) =  - {\left( g_1^D \right)^\alpha}_\beta {\bar D}_{R, \alpha }\Phi_1^\dag Q_L^\beta -
{\left( g_1^U \right)^\alpha}_\beta {\bar U}_{R, \alpha}{\tilde \Phi}_1^\dag Q_L^\beta \nonumber\\
&\qquad\qquad\qquad\;\: - {\left( g_2^D \right)^\alpha}_\beta {\bar D}_{R, \alpha } \Phi_2^\dag Q_L^\beta  - {\left( g_2^U
\right)^\alpha}_\beta {\bar U}_{R, \alpha}{\tilde \Phi}_2^\dag Q_L^\beta + {\rm h.c.},\nonumber \\
&L_{Y2}(S) =  - {\left( g_3^D \right)^\alpha}_\beta {\bar D}_{R, \alpha}S^\dag Q_L^\beta  -
{\left( g_3^U \right)^\alpha}_\beta {\bar U}_{R, \alpha }{\tilde S}^\dag Q_L^\beta  + {\rm h.c.}
\label{yukawas}
\end{eqnarray}
where ${\tilde H}_i = \varepsilon_{ij} H_j^*$ as usual 
 for all three scalar doublets $H=\Phi_{1,2},S$,  and $\alpha, \beta$ are flavour indices.

\subsection{Minimal flavour Violation}

The usual approach to suppressing FCNC in 2HDM is to introduce discrete symmetries that force for the Type I, $g_1^{D,U} = 0$, for Type II, $g_1^U = g_2^D = 0$.The type I can be enforced with  $\phi_1 \to  - \phi_1$, and the type II  with  $\phi_1 \to - \phi_1$, $d_R \to  - d_R$ \cite{reviews}. We use instead MFV \cite{Manohar:2006ga}  requiring  that there be only two flavour symmetry breaking matrices: $G^U$ which transforms as $(3_U,\bar{3}_Q)$ under the flavour group and $G^D$which transforms as $(3_D,\bar{3}_Q)$. The matrices appearing in Eq.~\ref{yukawas} become
\begin{eqnarray}
g_1^D=\eta_1^D G^D,\, g_2^D=\eta_2^D G^D,\, g_3^D=\eta_3^D G^D \nonumber\\
g_1^U=\eta_1^U G^U,\, g_2^U=\eta_2^U G^U,\, g_3^U=\eta_3^U G^U.
\end{eqnarray}
where $\eta_i^{D, U}$, $i = 1, 2, 3$, are complex scalars. The two types of two Higgs doublet model can be defined by 
\begin{itemize}
\item Type I: $\eta_1^D=\eta_1^U =0$ 
\item Type II: $\eta_1^U=\eta_2^D=0$
\end{itemize}

MFV is less restrictive than the discrete symmetries and allows quartic terms in the scalar potential that are odd in either of the doublets. In particular it allows the terms with coefficients $\nu_{4,5}$, $\omega_{4,5}$ and $\kappa_{1,2,3}$ in Eqs.~\ref{phisV}~and~\ref{3fields}.  It also allows the additional terms in the regular 2HDM sector,
\begin{eqnarray}
V^\prime(\Phi_1,\Phi_2) = \lambda_6 \left( \Phi_1^\dag \Phi_1 \right)\left( \Phi_1^\dag \Phi_2 \right)+\lambda_7 \left( \Phi_2^\dag \Phi_2 \right)\left( \Phi_1^\dag \Phi_2 \right) +{\rm h.c.}.
\end{eqnarray}
We do not include these two terms in our numerical studies.

\subsection{Custodial symmetry}

To impose custodial symmetry we follow the matrix formulation \cite{Pomarol:1993mu}. Scalar doublets are written as,
\begin{eqnarray}
&{M_{ab}} = \left( {{{\tilde \Phi }_a}, {\Phi _b}} \right) =
\begin{pmatrix}
{\phi _a^{0*}}&{\phi _b^ + }\\
{ - \phi _a^ - }&{\phi _b^0}
\end{pmatrix},\ 
a, b = 1,2,\\
&{{\cal S}^A} = \left( {{{\tilde S}^A}, {S^A}} \right) =
\begin{pmatrix}
{{S^{A0*}}}&{{S^{A + }}}\\
{ - {S^{A - }}}&{{S^{A0}}}
\end{pmatrix},
\end{eqnarray}
and the custodial symmetry is imposed by writing the scalar potential directly in terms of $O(4)$ invariants.

There are two methods proposed in the literature, 
\begin{itemize}
\item Using only ${M_{11}}$ and ${M_{22}}$. This results is all the $\lambda_i$ being real and 
\begin{eqnarray}
{\kappa_2} = {\kappa _3},\ 2{\nu _3} = {\nu _2},\ {\nu _4} = \nu _5^*,\ 2{\omega _3} = {\omega_2},\ {\omega _4} = \omega _5^*,\ \lambda_4 = \lambda_5.
\label{met1}
\end{eqnarray}

\item Using only ${M_{12}}$ results instead in
\begin{eqnarray}
&&{\nu_2}={\omega_2}={\kappa_3} = {\kappa _3^\star},\  \kappa_2=2\nu_2,\  \nu_3=\omega_3^\star,\ \nonumber \\
 &&\lambda_6 = \lambda_7,\ \lambda_1=\lambda_2=\lambda_3,\ m_{11}^2=m_{22}^2.
 \label{met2}
\end{eqnarray}
For the vacuum to be invariant as well one needs $v_1^\star=v_2$ which is too restrictive so we will only use the first method.
\end{itemize}

Imposing custodial symmetry results in  ${\rm{\Delta}} \rho = 0$ up weak corrections as can be easily verified.
It also results in $m_{H^\pm}=m_A$  and in $m_{S^\pm}=m_{S_I^0}$. It has been pointed out before that it is also possible to satisfy ${\rm{\Delta}} \rho = 0$ with $m_{H^\pm}=m_H$ \cite{Gerard:2007kn,Cervero:2012cx} and in $m_{S^\pm}=m_{S_R^0}$ \cite{Burgess:2009wm}, and that this follows from `twisted' custodial symmetry.

The results in Eq.~\ref{met1} can be compared with those obtained in the SM plus MW case,
\begin{eqnarray}
2\tilde{\lambda}_3=\tilde{\lambda}_2,\, 2\tilde{\lambda}_6=2\tilde{\lambda}_7=\tilde{\lambda}_{11},\, \tilde{\lambda}_9=\tilde{\lambda}_{10},\, \tilde{\lambda}_4=\tilde{\lambda}_5^\star.
\label{custsym}
\end{eqnarray}

\subsection{Unitarity constraints}

In this section we consider high energy two-to-two scalar scattering to constrain the strength of the self interactions with the requirement of perturbative unitarity. Although the potential is renormalizable, the tree-level scattering amplitudes approach a constant at high energy that is proportional to the quartic couplings. Perturbative unitarity then constrains their size in a manner entirely analogous to the unitarity bound on the SM Higgs boson mass \cite{Lee:1977eg} and generalizations \cite{Kanemura:1993hm, Horejsi:2005da,Ginzburg:2005dt}. We will consider scattering of all the scalar particles that appear in the model at energies much larger than their masses. The strongest limits on the couplings are obtained by considering scattering of two particle states of definite colour and $I=0$.  In this context, $I=0$ is the singlet of the approximate $O(4)$ symmetry. 

\begin{figure}[htb]
\includegraphics[width=1\textwidth]{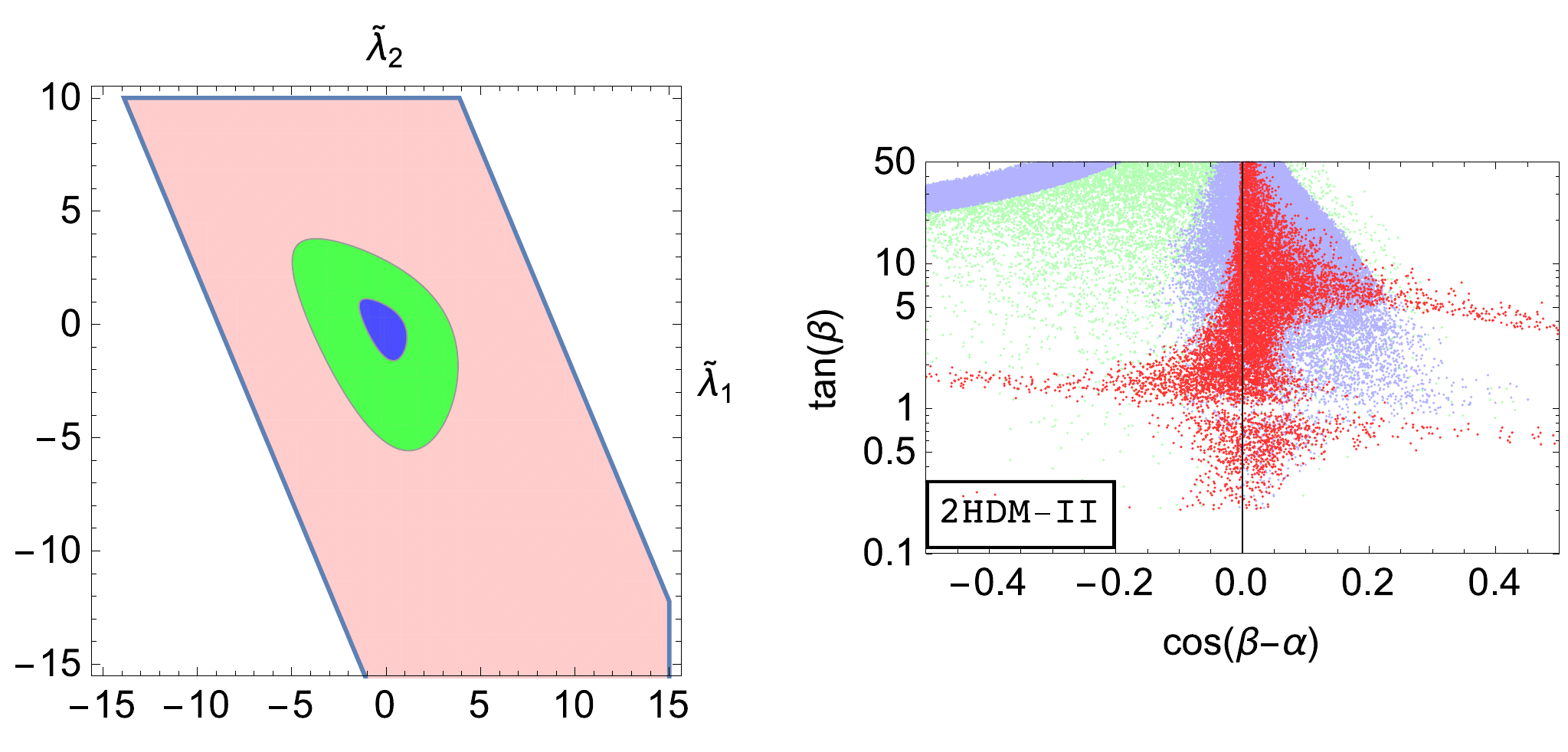}
\caption{The left panel shows the region in the $\tilde\lambda_1-\tilde\lambda_2$  plane that satisfies the unitarity constraint in the SM plus MW at 1~TeV in red. Including unitarity constraints on the running of $m_h$ up to 100~TeV in green and up to $10^{10}$~GeV in blue. The right panel shows a 
comparison of unitarity constraints (red points) to  $1\sigma$ constraints  from $h\to gg$ and $h \to \gamma\gamma$ in the 2HDM-II (blue points) and the 2HDM plus a colour octet (green).}
\label{f:unit} 
\end{figure}

We begin by showing some of the resulting constraints for the SM plus MW case and comparing tree level unitarity with the improvement one can get by allowing the couplings to run. \cite{Marciano:1989ns} A further improvement is achieved by including the modifications to the running of the SM quartic coupling (or Higgs mass).~\cite{He:2013tla} This can be seen on the left panel of Figure~\ref{f:unit}.

We next extend unitarity constraints to the 2HDM plus MW case, with the additional requirement of the known conditions for having a positive definite Higgs potential with a $Z_2$ symmetry \cite{Deshpande:1977rw},
\begin{equation}
\lambda_1 > 0, \quad
\lambda_2 > 0, \quad
\lambda_3 > - \sqrt{\lambda_1 \lambda_2},\quad
\lambda_3 + \lambda_4 \pm \lambda_5 > - \sqrt{\lambda_1 \lambda_2}.
\label{stabvac}
\end{equation}
We will always identify the lightest neutral scalar $h$ with the 125.6~GeV state found at LHC \cite{Aad:2012tfa,Chatrchyan:2012ufa}. The heavy scalar is allowed to have a mass in the range $600\leq m_H \leq 900$~GeV and the pseudoscalar and charged scalars in the range $400\leq m_A=m_{H^\pm} \leq 1300$~GeV. For the case of the 2HDM, the the two-to-two scattering matrix for neutral colour singlets is $14\times14$ and can be diagonalized exactly \cite{Kanemura:1993hm}. When the colour octet is added the matrix becomes $18\times 18$ and we diagonalize it numerically.  
Unitarity constraints are obtained again from the $J=0$ partial wave. 
We show one of the more interesting projections of these constraints on the right panel of Figure~\ref{f:unit}.

The region allowed by tree-level unitarity for the sector of the potential  that couples the colour singlets and octets shows approximate correlations of the form $|2\nu_1+\nu_2|\lsim 14$, $|2\omega_1+\omega_2|\lsim 15$ and $|2\kappa_1+\kappa_2|\lsim 11$ \cite{Cheng:2016tlc}. This is also the correlation observed in SM + MW for $\tilde\lambda_1-\tilde\lambda_2$ that is shown in Figure~\ref{f:unit}.

\subsection{Tree-level Higgs decay}

\begin{figure}[htb]
\includegraphics[width=.9\textwidth]{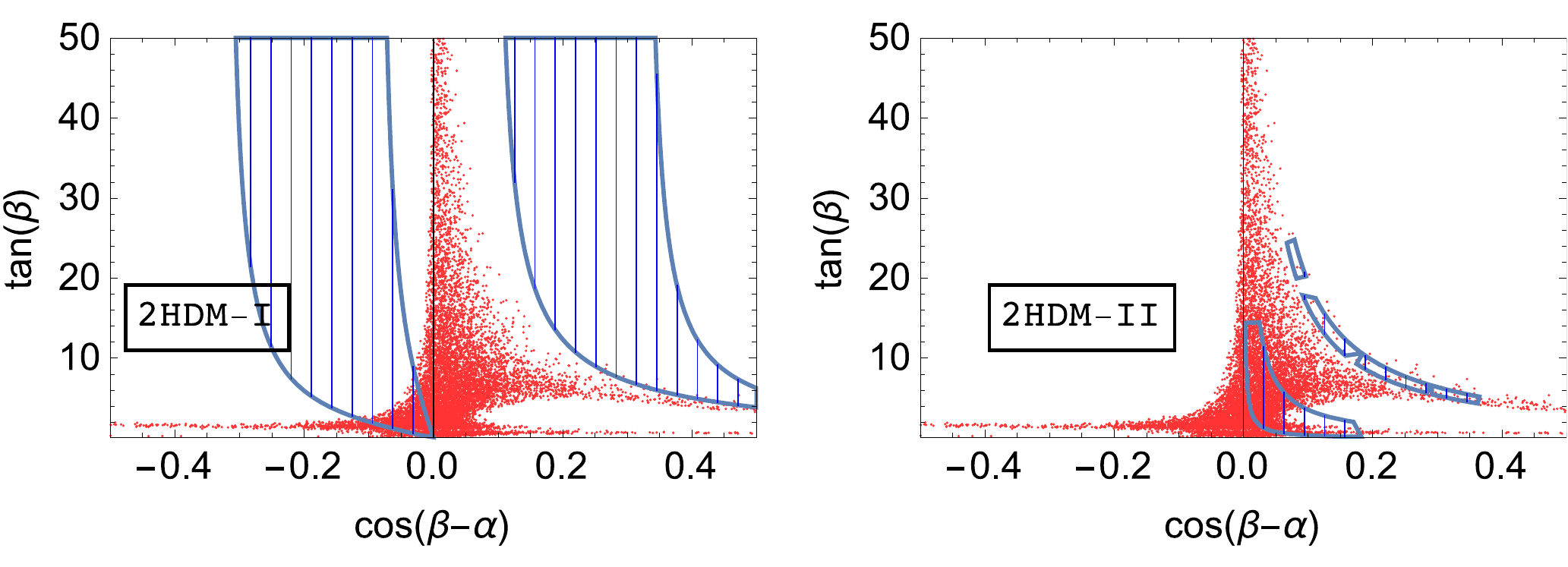}
\caption{
\label{tree-con} Constraints on the $\cos(\beta-\alpha)-\tan\beta$ plane arising from LHC fits to $\kappa_t$, $\kappa_b$ and $\kappa_\tau$.Left for 2HDM-I at $2.2\sigma$ (dashed blue) and right for 2HDM-II  at $2\sigma$.  The red  area is that allowed by tree-level unitarity.}
\end{figure}

The tree-level Higgs couplings to $t\bar{t}$, $b\bar{b}$ and $\tau^+\tau^-$ as well as  to $WW$ and $ZZ$ already constrain the parameter space of the 2HDM requiring it to be close to the SM \cite{Barroso:2013zxa,Ferreira:2013qua,Haber:2015pua}. We illustrate the following constraints in Figure~\ref{tree-con} as per the ATLAS-CMS combination of data for the case where BSM physics is allowed in loops and in decays \cite{atlas-cms},
\begin{eqnarray}
\kappa_b = 0.60^{+0.18}_{-0.18},\quad
\kappa_\tau = 0.88^{+0.13}_{-0.12},\quad
\kappa_t = 1.43^{+0.23}_{-0.22}.
\end{eqnarray}

On the left panel we look at the type-I 2HDM and show the region allowed at the $2.2\sigma$ level by the ATLAS-CMS combination (dashed blue) superimposed on the region allowed by tree-level unitarity (red dots). 
We use $2.2\sigma$ because the 2HDM-I is ruled out by this data at the $2\sigma$ level  
due to the conflicting requirements of an enhanced top-quark coupling and a reduced $b$-quark coupling. On the right  panel we show the type-II 2HDM at $2\sigma$. In this case the strongest constraint arises from the tau-lepton couplings \cite{Cheng:2016tlc}.

\subsection{Direct bounds on the colour octet}

One would expect that the LHC can rule out additional light colour scalars from their non-observation. It turns out however that the existing bounds are not very restrictive.

The basic constraints arise from decays into two jets or a $ t\bar{t}$ pair. CMS limits on a colour-octet scalar $S^0$ from dijet final state quote $M_S < 3.1$~TeV \cite{Khachatryan:2015dcf}. However, this is a for a model with production cross-section a few thousand times larger. Similarly, bounds on $Z^\prime$ resonances decaying to  $ t\bar{t}$ pairs \cite{Chatrchyan:2012yca} can be interpreted as posing no significant constraint for these scalars. It is known \cite{Manohar:2006ga,Gresham:2007ri} that the cross sections for producing pairs of coloured scalars are larger than those for single scalar production for much of the parameter space. The relevant constraints would then be dijet pairs and four top-quarks. But again the models that have been studied have much larger production cross-sections than MW and effectively there are no direct constraints from LHC yet.

\subsection{One loop constraints}

The additional contribution to the Higgs boson production due to the octet-scalar loops  in the limit of very heavy quarks and colour scalars in the loop can be written as  \cite{Manohar:2006ga,He:2011ti}
\begin{eqnarray}
{\cal L} = (\sqrt{2} G_F)^{1/2} \frac{\alpha_s}{12\pi}\
G^A_{\mu\nu}G^{A\mu\nu} h \left(n_{hf} +\frac{v^2}{m^2_S}\frac{3}{8}(2\tilde\lambda_1+\tilde\lambda_2)\right)
\label{effhgg}
\end{eqnarray}
where $n_{hf}$ is the number of heavy quark flavours, one in the case of SM3 and three in the case of SM4. This result shows the important role that additional colour scalars can play in the effective one-loop couplings of the Higgs.

\begin{figure}[thb]
\includegraphics[width=1\textwidth]{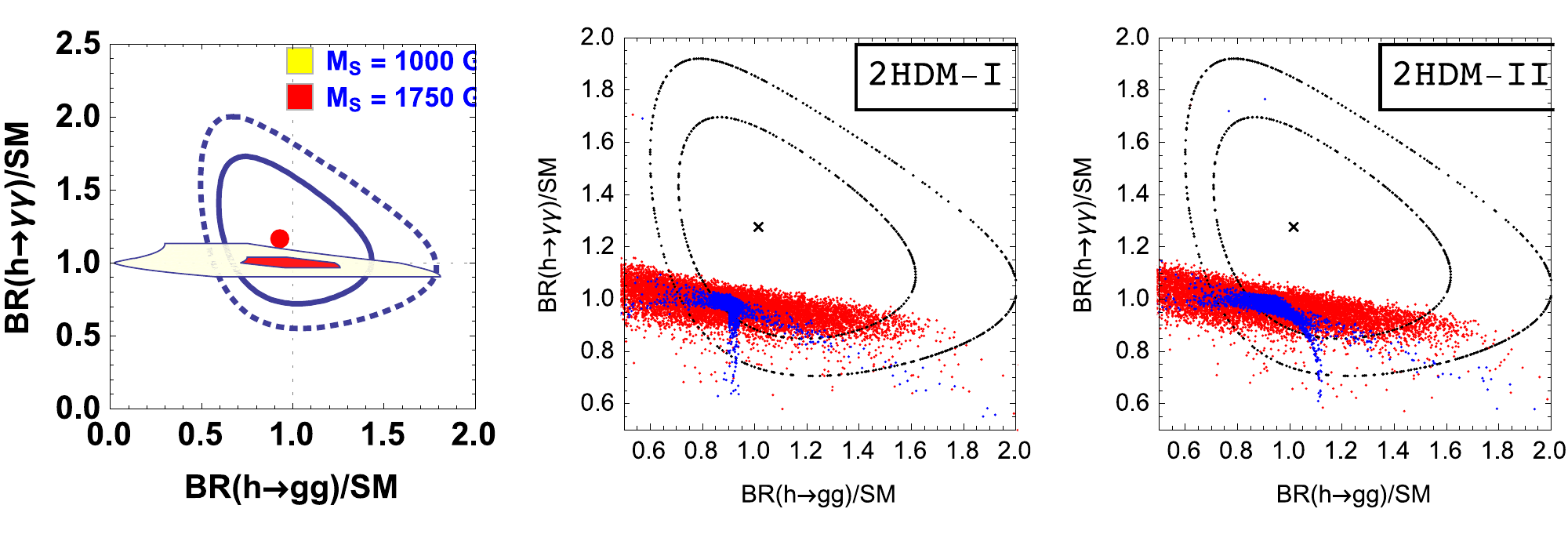}
\caption{
\label{f:hloop} Best fit to $BR(h\to \gamma\gamma)$ vs $BR(h\to gg)$\cite{Giardino:2013bma}: the red dot (black x) is the best fit, the solid and dashed curves show the $1\sigma$ and $2\sigma$ allowed regions respectively. In the left panel we have superimposed the range of predictions in the SM + MW for two values of $M_S$ and values of $\tilde\lambda_{1,2}$ spanning the parameter space allowed by tree-level unitarity. On the right panel we superimpose  the parameter space that satisfies the unitarity constraints for the 2HDM (blue points) and for the 2HDM + MW (red points). }
\end{figure}

Next we present the points allowed by tree-level unitarity in a $h\to gg$ vs $h \to \gamma\gamma$ plot in Figure~\ref{f:hloop}. The black contours are taken from {\it The universal Higgs fit}~\cite{Giardino:2013bma}.\footnote{We thank Kristjan Kannike who provided us with these fits.}  The SM point is, of course, (1,1). On the left panel we have overlaid the region allowed by tree level unitarity of the SM plus MW model for two values of $M_S$, 1~TeV and 1.75~TeV. On the right panel we have overlaid the blue regions which are the points allowed by unitarity for the 2HDM parameter space, and the red regions corresponding to the 2HDM augmented by the colour-octet.
The figure illustrates how the loop induced Higgs decays are at present the best channels to constrain a Manohar-Wise type colour-octet. The colour-octet also extends the region which can be explained with a 2HDM mostly in the direction of a larger $BR(h\to gg)$.

\begin{figure}[thb]
\includegraphics[width=.8\textwidth]{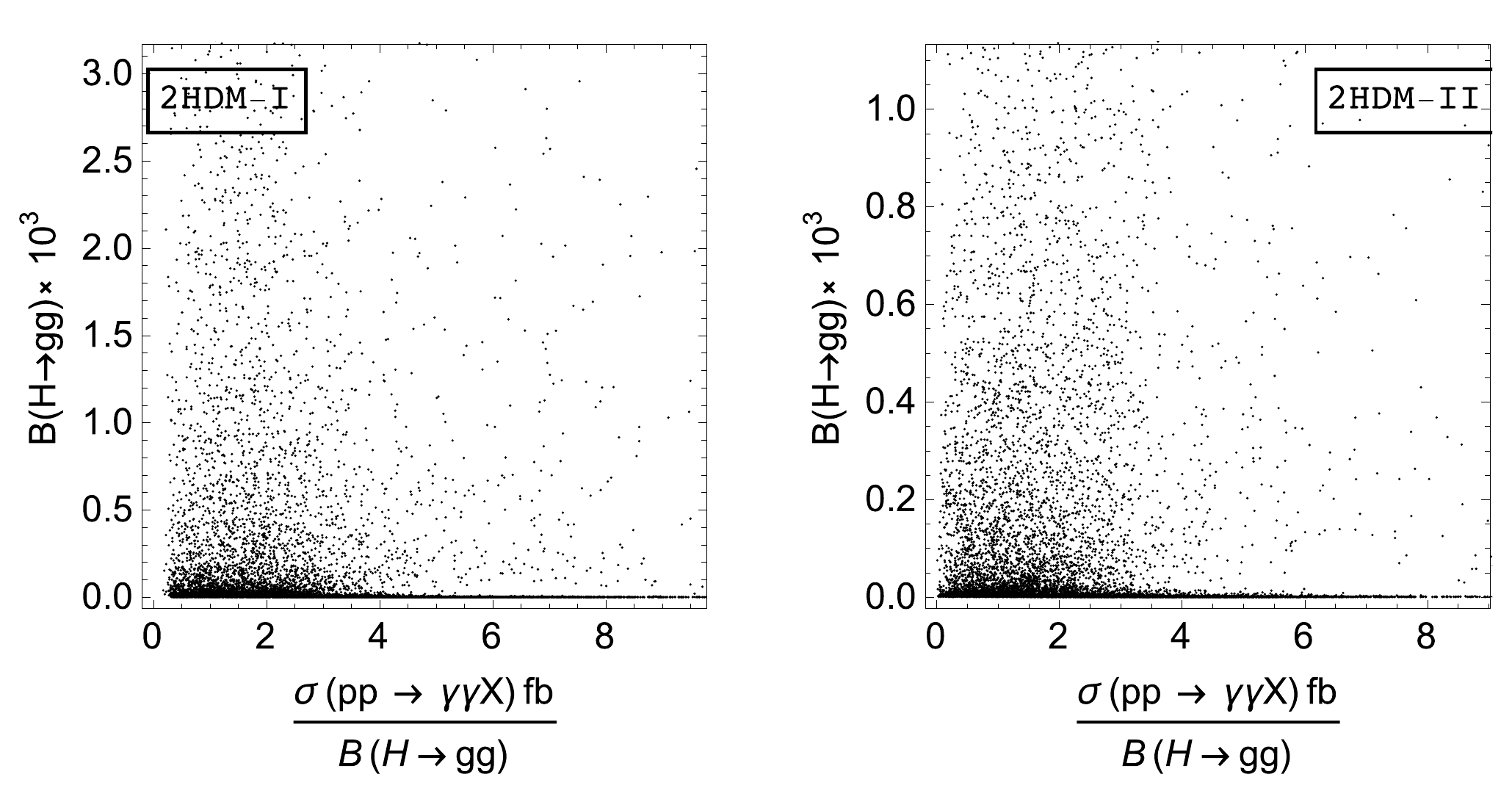}
\caption{
\label{f:dipho} Cross-section $\sigma(pp\to\gamma\gamma X)$ normalised to $B(H\to gg)$  through a 750~GeV $H^0$ as a function of  $B(H\to gg)$ for points allowed by both unitarity and $h\to gg$, $h\to \gamma\gamma$ at $1\sigma$.}
\end{figure}

Recent interest in a possible 750~GeV di-photon resonance \cite{CMS:2015dxe,ATLAS750}, compels us to explore the possibility of it being the $H^0$ in a 2HDM extended with a colour-octet. Other papers have used a colour-octet in this context recently \cite{Ding:2016ldt,Cao:2015twy}. Figure~\ref{f:dipho} shows this is not possible. The observed signal interpreted as a resonance requires  a cross-section $\sigma(pp\to\gamma\gamma X)$ in the vicinity of 10fb\cite{Altmannshofer:2015xfo}. The largest $B(H\to gg)$ that can be obtained are about three orders of magnitude too small to reach the necessary cross-section. 

\section{Conclusions}

We discussed Higgs phenomenology of a scalar sector augmented with a new multiplet of colour octet scalars as in the Manohar-Wise model which is motivated by MFV both for a one Higgs doublet model (SM) and a two Higgs doublet model (type I and II). Starting from the most general renormalizable scalar potential we have reduced the number of allowed terms with the usual theoretical requirements of minimal flavor violation and custodial symmetry. 
We considered constraints on the masses and  parameters of the model from unitarity and vacuum stability and saw that the additional colour scalars are very difficult to rule out or to observe directly. 

The measured $h g g$ and $h\gamma\gamma$ couplings are in agreement with the SM, but there is still room for new physics at the ~50\% level. They constitute one of the best places to constrain a MW extension. 
For SM plus MW we found that the Higgs one-loop effective couplings place constraints on the model that fall between those from tree level unitarity and those from RGI unitarity. 
We found that the existence of such colour scalars would effectively remove the constraints on the top Yukawa coupling arising from these couplings.

We  confronted the model with available LHC results in the form of fitted couplings of the Higgs boson which we identify with the lightest scalar in the 2HDM. After collecting constraints on the parameters of the 2HDM from tree-level Higgs couplings we constrain the new sector couplings to the colour-octet using a current fit on the one loop $h\to \gamma\gamma$ and $h\to gg$ couplings.

Finally we predict the one loop couplings of the heavier neutral scalar $H\to \gamma\gamma$ and $H\to gg$ using the points in parameter space that satisfy all our constraints. We find that this cannot be the 750~GeV di-photon resonance that might have been seen at LHC.

This research was supported in part by the DOE under contract number DE-SC0009974. We thank  Harald Fritzsch for the opportunity to participate in a very productive meeting.


\begin{thebibliography}{999}        
\bibitem{Aad:2012tfa} 
  G.~Aad {\it et al.}  [ATLAS Collaboration],
  Phys.\ Lett.\ B {\bf 716}, 1 (2012)
  [arXiv:1207.7214 [hep-ex]].
  
\bibitem{Chatrchyan:2012ufa} 
  S.~Chatrchyan {\it et al.}  [CMS Collaboration],
  Phys.\ Lett.\ B {\bf 716}, 30 (2012)
  [arXiv:1207.7235 [hep-ex]].
  
\bibitem{CMS:2015dxe} 
  CMS Collaboration [CMS Collaboration],
  collisions at 13TeV,''
  CMS-PAS-EXO-15-004.

 \bibitem{ATLAS750} 
  The ATLAS collaboration,
  ATLAS-CONF-2015-081.
  
\bibitem{Glashow:1976nt} 
  S.~L.~Glashow and S.~Weinberg,
  Phys.\ Rev.\ D {\bf 15}, 1958 (1977).
  doi:10.1103/PhysRevD.15.1958
  
\bibitem{Chivukula:1987py}
  R.~S.~Chivukula and H.~Georgi,
  Phys.\ Lett.\ B {\bf 188} (1987) 99.
  
\bibitem{D'Ambrosio:2002ex}
  G.~D'Ambrosio, G.~F.~Giudice, G.~Isidori and A.~Strumia,
  Nucl.\ Phys.\ B {\bf 645} (2002) 155
  [hep-ph/0207036].

  
\bibitem{Manohar:2006ga} 
  A.~V.~Manohar and M.~B.~Wise,
  Phys.\ Rev.\ D {\bf 74}, 035009 (2006)
  [hep-ph/0606172].

\bibitem{reviews}
See for example, 
  J.~F.~Gunion, H.~E.~Haber, G.~L.~Kane and S.~Dawson,
  Front.\ Phys.\  {\bf 80}, 1 (2000).
  G.~C.~Branco, P.~M.~Ferreira, L.~Lavoura, M.~N.~Rebelo, M.~Sher and J.~P.~Silva,
  Phys.\ Rept.\  {\bf 516}, 1 (2012)
  [arXiv:1106.0034 [hep-ph]].
and references therein.

\bibitem{Arnold:2009ay} 
  J.~M.~Arnold, M.~Pospelov, M.~Trott and M.~B.~Wise,
  JHEP {\bf 1001}, 073 (2010)
  doi:10.1007/JHEP01(2010)073
  [arXiv:0911.2225 [hep-ph]].
HEP :: Search ::  Help ::  Terms of use ::  Privacy policy 
Powered by Invenio v1.1.2+ 
Problems/Questions to feedback@inspirehep.net 

\bibitem{Perez:2016qbo} 
  P.~Fileviez Perez and C.~Murgui,
  arXiv:1604.03377 [hep-ph].
  
\bibitem{Bertolini:2013vta} 
  S.~Bertolini, L.~Di Luzio and M.~Malinsky,
  Phys.\ Rev.\ D {\bf 87}, no. 8, 085020 (2013)
  doi:10.1103/PhysRevD.87.085020
  [arXiv:1302.3401 [hep-ph]].

\bibitem{Hill:1991at} 
  C.~T.~Hill,
  Phys.\ Lett.\ B {\bf 266}, 419 (1991).
  doi:10.1016/0370-2693(91)91061-Y
  
\bibitem{Farhi:1980xs} 
  E.~Farhi and L.~Susskind,
  Phys.\ Rept.\  {\bf 74}, 277 (1981).
  doi:10.1016/0370-1573(81)90173-3
  
\bibitem{He:2011ti} 
  X.~-G.~He and G.~Valencia,
  Phys.\ Lett.\ B {\bf 707}, 381 (2012)
  [arXiv:1108.0222 [hep-ph]].
  
\bibitem{He:2013tia} 
  X.~G.~He, Y.~Tang and G.~Valencia,
  Phys.\ Rev.\ D {\bf 88}, 033005 (2013)
  doi:10.1103/PhysRevD.88.033005
  [arXiv:1305.5420 [hep-ph]].
  
\bibitem{He:2011ws} 
  X.~-G.~He, G.~Valencia and H.~Yokoya,
  JHEP {\bf 1112}, 030 (2011)
  [arXiv:1110.2588 [hep-ph]].

\bibitem{Burgess:2009wm} 
  C.~P.~Burgess, M.~Trott and S.~Zuberi,
  JHEP {\bf 0909}, 082 (2009)
  [arXiv:0907.2696 [hep-ph]].
  
\bibitem{Carpenter:2011yj} 
  L.~M.~Carpenter and S.~Mantry,
  Phys.\ Lett.\ B {\bf 703}, 479 (2011)
  [arXiv:1104.5528 [hep-ph]].
  
\bibitem{Enkhbat:2011qz} 
  T.~Enkhbat, X.~-G.~He, Y.~Mimura and H.~Yokoya,
  JHEP {\bf 1202}, 058 (2012)
  [arXiv:1105.2699 [hep-ph]].

\bibitem{Dobrescu:2011aa} 
  B.~A.~Dobrescu, G.~D.~Kribs and A.~Martin,
  Phys.\ Rev.\ D {\bf 85}, 074031 (2012)
  [arXiv:1112.2208 [hep-ph]].
  
\bibitem{Bai:2011aa} 
  Y.~Bai, J.~Fan and J.~L.~Hewett,
  JHEP {\bf 1208}, 014 (2012)
  [arXiv:1112.1964 [hep-ph]].
 
\bibitem{Arnold:2011ra} 
  J.~M.~Arnold and B.~Fornal,
  Phys.\ Rev.\ D {\bf 85}, 055020 (2012)
  [arXiv:1112.0003 [hep-ph]].
 
\bibitem{Cacciapaglia:2012wb} 
  G.~Cacciapaglia, A.~Deandrea, G.~D.~La Rochelle and J.~-B.~Flament,
  arXiv:1210.8120 [hep-ph].
 
\bibitem{Dorsner:2012pp} 
  I.~Dorsner, S.~Fajfer, A.~Greljo and J.~F.~Kamenik,
  arXiv:1208.1266 [hep-ph].
  
\bibitem{Kribs:2012kz} 
  G.~D.~Kribs and A.~Martin,
  arXiv:1207.4496 [hep-ph].
    
\bibitem{Reece:2012gi} 
  M.~Reece,
  arXiv:1208.1765 [hep-ph].
  
\bibitem{Cao:2013wqa} 
  J.~Cao, P.~Wan, J.~M.~Yang and J.~Zhu,
  arXiv:1303.2426 [hep-ph].

\bibitem{He:2013tla} 
  X.~G.~He, H.~Phoon, Y.~Tang and G.~Valencia,
  JHEP {\bf 1305}, 026 (2013)
  doi:10.1007/JHEP05(2013)026
  [arXiv:1303.4848 [hep-ph]].


\bibitem{Cheng:2015lsa} 
  X.~D.~Cheng, X.~Q.~Li, Y.~D.~Yang and X.~Zhang,
  J.\ Phys.\ G {\bf 42}, no. 12, 125005 (2015)
  doi:10.1088/0954-3899/42/12/125005
  [arXiv:1504.00839 [hep-ph]].

\bibitem{Buttazzo:2014bka} 
  D.~Buttazzo,
  arXiv:1403.6535 [hep-ph].

\bibitem{He:2014xla} 
  X.~G.~He, G.~N.~Li and Y.~J.~Zheng,
  Int.\ J.\ Mod.\ Phys.\ A {\bf 30}, no. 25, 1550156 (2015)
  doi:10.1142/S0217751X15501560
  [arXiv:1501.00012 [hep-ph]].
\bibitem{Yue:2014tya} 
  J.~Yue,
  Phys.\ Lett.\ B {\bf 744}, 131 (2015)
  doi:10.1016/j.physletb.2015.03.044
  [arXiv:1410.2701 [hep-ph]].


\bibitem{Pomarol:1993mu} 
  A.~Pomarol and R.~Vega,
  Nucl.\ Phys.\ B {\bf 413}, 3 (1994)
  doi:10.1016/0550-3213(94)90611-4
  [hep-ph/9305272].
  
  
\bibitem{Lee:1977eg} 
  B.~W.~Lee, C.~Quigg and H.~B.~Thacker,
  Phys.\ Rev.\ D {\bf 16}, 1519 (1977).
  
\bibitem{Kanemura:1993hm} 
  S.~Kanemura, T.~Kubota and E.~Takasugi,
  Phys.\ Lett.\ B {\bf 313}, 155 (1993)
  [hep-ph/9303263].

\bibitem{Horejsi:2005da} 
  J.~Horejsi and M.~Kladiva
  Eur.\ Phys.\ J.\ C {\bf 46}, 81 (2006)
  doi:10.1140/epjc/s2006-02472-3
  [hep-ph/0510154].

\bibitem{Ginzburg:2005dt} 
  I.~F.~Ginzburg and I.~P.~Ivanov,
  Phys.\ Rev.\ D {\bf 72}, 115010 (2005)
  doi:10.1103/PhysRevD.72.115010
  [hep-ph/0508020].
  
   
\bibitem{Marciano:1989ns} 
  W.~J.~Marciano, G.~Valencia and S.~Willenbrock,
  Phys.\ Rev.\ D {\bf 40}, 1725 (1989).
  doi:10.1103/PhysRevD.40.1725

\bibitem{Gerard:2007kn} 
  J.-M.~Gerard and M.~Herquet,
  Phys.\ Rev.\ Lett.\  {\bf 98}, 251802 (2007)
  doi:10.1103/PhysRevLett.98.251802
  [hep-ph/0703051 [HEP-PH]].

\bibitem{Cervero:2012cx} 
  E.~Cerver— and J.~M.~GŽrard,
ÊÊPhys.\ Lett.\ B {\bf 712}, 255 (2012)
ÊÊdoi:10.1016/j.physletb.2012.05.010
ÊÊ[arXiv:1202.1973 [hep-ph]].
ÊÊ
  
\bibitem{Deshpande:1977rw} 
  N.~G.~Deshpande and E.~Ma,
  Phys.\ Rev.\ D {\bf 18}, 2574 (1978).
  doi:10.1103/PhysRevD.18.2574

\bibitem{Cheng:2016tlc} 
  L.~Cheng and G.~Valencia,
  arXiv:1606.01298 [hep-ph].

\bibitem{Barroso:2013zxa} 
  A.~Barroso, P.~M.~Ferreira, R.~Santos, M.~Sher and J.~P.~Silva,
  arXiv:1304.5225 [hep-ph].

\bibitem{Ferreira:2013qua} 
  P.~M.~Ferreira, R.~Santos, M.~Sher and J.~P.~Silva,
  arXiv:1305.4587 [hep-ph].
  
\bibitem{Haber:2015pua} 
  H.~E.~Haber and O.~Stal,
  Eur.\ Phys.\ J.\ C {\bf 75}, no. 10, 491 (2015)
  doi:10.1140/epjc/s10052-015-3697-x
  [arXiv:1507.04281 [hep-ph]].

  
\bibitem{atlas-cms} 
  The ATLAS and CMS Collaborations,
  ATLAS-CONF-2015-044.

\bibitem{Khachatryan:2015dcf} 
  V.~Khachatryan {\it et al.} [CMS Collaboration],
  Phys.\ Rev.\ Lett.\  {\bf 116}, no. 7, 071801 (2016)
  doi:10.1103/PhysRevLett.116.071801
  [arXiv:1512.01224 [hep-ex]].
 
\bibitem{Chatrchyan:2012yca} 
  S.~Chatrchyan {\it et al.} [CMS Collaboration],
  Phys.\ Rev.\ D {\bf 87}, no. 7, 072002 (2013)
  doi:10.1103/PhysRevD.87.072002
  [arXiv:1211.3338 [hep-ex]].

\bibitem{Gresham:2007ri} 
  M.~I.~Gresham and M.~B.~Wise,
  Phys.\ Rev.\ D {\bf 76}, 075003 (2007)
  [arXiv:0706.0909 [hep-ph]].

\bibitem{Giardino:2013bma} 
  P.~P.~Giardino, K.~Kannike, I.~Masina, M.~Raidal and A.~Strumia,
  JHEP {\bf 1405}, 046 (2014)
  doi:10.1007/JHEP05(2014)046
  [arXiv:1303.3570 [hep-ph]].

  
 
\bibitem{Ding:2016ldt} 
  R.~Ding, Z.~L.~Han, Y.~Liao and X.~D.~Ma,
  Eur.\ Phys.\ J.\ C {\bf 76}, no. 4, 204 (2016)
  doi:10.1140/epjc/s10052-016-4052-6
  [arXiv:1601.02714 [hep-ph]].
\bibitem{Cao:2015twy} 
  J.~Cao, C.~Han, L.~Shang, W.~Su, J.~M.~Yang and Y.~Zhang,
  Phys.\ Lett.\ B {\bf 755}, 456 (2016)
  doi:10.1016/j.physletb.2016.02.045
  [arXiv:1512.06728 [hep-ph]].
  
\bibitem{Altmannshofer:2015xfo} 
  W.~Altmannshofer, J.~Galloway, S.~Gori, A.~L.~Kagan, A.~Martin and J.~Zupan,
  Phys.\ Rev.\ D {\bf 93}, no. 9, 095015 (2016)
  doi:10.1103/PhysRevD.93.095015
  [arXiv:1512.07616 [hep-ph]].

 
\end{thebibliography}
\end{document}